\documentclass{ifacconf}

\usepackage{graphicx}      
\usepackage{natbib}        
\usepackage{amsmath,amsfonts,mathrsfs,amssymb,dsfont} 
\usepackage{tikz}
\usepackage{tabu}
\usepackage{mathtools}
\mathtoolsset{showonlyrefs}

\begin{document}
\begin{frontmatter}

\title{Learning without Recall: \\ A Case for Log-Linear Learning\thanksref{footnoteinfo}} 

\thanks[footnoteinfo]{This work was supported by ARO MURI W911NF-12-1-0509.}

\author{Mohammad Amin Rahimian \&} 
\author[Second]{Ali Jadbabaie} 

\address[Second]{Department of Electrical and Systems Engineering, University of Pennsylvania, Philadelphia, PA 19104-6228 USA \\ (e-mail: jadbabai@seas.upenn.edu)}

\begin{abstract}                
We analyze a model of learning and belief formation in networks in which  agents follow Bayes rule yet they do not recall their history of past observations and cannot reason about how other agents' beliefs are formed. They do so by making rational inferences about their observations which include a sequence of independent and identically distributed private signals as well as the beliefs of their neighboring agents at each time. Fully rational agents would successively apply Bayes rule to the entire history of observations. This leads to forebodingly complex inferences due to lack of knowledge about the global network structure that causes those observations. To address these complexities, we consider a ``\emph{Learning without Recall}'' model, which in addition to providing a tractable framework for analyzing the behavior of rational agents in social networks, can also provide a behavioral foundation for the variety of non-Bayesian update rules in the literature.  We present the implications of various choices for  time-varying priors of such agents and how this choice affects learning and its rate.
\end{abstract}


\end{frontmatter}

\section{Introduction \& Background}
Agents exchange beliefs in social networks to benefit from each other's opinions and private information in trying to learn an unknown state of the world. Rational agents in a social network would apply Bayes rule successively to their observations at each step, which include not only their private signals but also the beliefs communicated by their neighbors. However, such repeated applications of Bayes rule in networks become very complex, especially if the agents are unaware of the global network structure. This is due to the fact that the agents at each step should use their local data that is increasing with time, and make very complex inferences about possible signal structures leading to their observations. Indeed, tractable modeling and analysis of rational behavior in networks is an important problem in network economics and have attracted much attention, \cite{acemoglu2011bayesian,mueller2013general,MosselSlyTamuz14}.








To avoid the complexities of fully rational inference, a variety of non-Bayesian update rules have been proposed that  rely  on the seminal work of \cite{DeGrootModel} in linear opinion pooling, where agents update their opinions to a convex combination of their neighbors' beliefs and the coefficients correspond to the level of confidence that each agent puts in each of her neighbors. More recently,  \cite{Jadbabaie2012210,JaMoTa13} consider a variation of this model for streaming observations, where in addition to the neighboring beliefs the agents also receive private signals.  Other forms of non-Bayesian rules were studied by \cite{Bala01071998} who consider a variation of observational learning in which agents observe the action and pay-offs of their neighbors and make rational inferences about these action/pay-off correspondence together with the choices made by their neighbors, but ignore the fact that their neighbors are themselves learning from their own observations. More recently, \cite{eyster2010naive} consider models of autarkic play where players at each generation observe their predecessor but na\"{i}vely think that any predecessor's action relies solely on that player's private information, thus ignoring the possibility that successive generations are learning from each other.


A chief contribution of this paper is establishing a behavioral foundation for the existing non-Bayesian updates in the literature where the belief update rule has a log-linear structure. This paper addresses the question of how  one can limit the information requirement of a Bayesian update and still ensure consensus or learning for the agents.  Some of the non-Bayesian update rules have the property that they resemble the replication of a first step of a Bayesian update from a common prior, and the aim here is to formalize such a setup. For instance \cite{PersuasionBias} interpret the weights in the DeGroot model as those assigned initially by rational agents to the noisy opinions of their neighbors based on their perceived precision. However, by repeatedly applying the same weights over and over again, the agents ignore the need to update these weights with the increasing information. 

To this end, we propose the so-called \emph{Learning without Recall} model as a belief formation and update  rule for \emph{Rational but Memoryless} agents. We show how such a scheme can provide convergence and learning for all agents in a strongly connected social network so long as the truth is identifiable through the aggregate observations of the agents across entire network. This is of particular interest, when the agents cannot distinguish the truth based solely on their private observations, and yet together they learn.

\section{The Learning without Recall Model}

\paragraph*{\bf Notation.}  Throughout the paper, $\mathbb{N}_0 = \{0\}\cup\mathbb{N}$ is the set of positive integers and zero. Boldface letters denote random variables, the identity matrix is denoted by $I$, and $\left\lVert \mathord{\cdot} \right\rVert$ is a vector norm. We use Greek letters to denote certain variables of interest such as: agents beliefs ($\mu$), $\log$-ratio of beliefs ($\phi$),  $\log$-likelihood ratio of signals ($\lambda$), and $\log$-ratio of initial prior beliefs ($\psi$). We consider a network of $n$ agents labeled by $[n]:=\{1,2,\ldots,n\}$, that interact according to a directed graph $\mathcal{G} = ([n],\mathcal{E})$, where $\mathcal{E} \subset [n] \times [n]$ is the set of directed edges. Each agent is labeled by a unique element of the set $[n]$. $\mathcal{N}(i) = \{j \in [n]; (j,i) \in \mathcal{E}\}$ is  the neighborhood of agent $i$, which is the set of all agents whose beliefs can be observed by agent $i$. We let $\deg(i) = \mid\mathcal{N}(i)\mid$ be the degree of node $i$ corresponding to the number of agent $i$'s neighbors. 
\vspace*{-12pt}
\paragraph*{{\bf Signals and Environment.}} We denote by $\Theta$ the finite set of states of the world. Also, $\Delta\Theta$ represents the space of all probability measures on the set $\Theta$. Each agent's goal is to decide amongst the finitely many possibilities in the state space $\Theta$. A random variable $\boldsymbol{\theta}$ is chosen randomly from $\Theta$ by nature and according to the probability measure $\nu(\mathord{\cdot}) \in \Delta\Theta$, which satisfies $\nu(\check{\theta}) > 0,\forall \check{\theta} \in \Theta$. For each agent $i$, there exists a finite signal space denoted by $\mathcal{S}_i$, and given $\boldsymbol{\theta}$, $\ell_i(\mathord{\cdot}\mid\boldsymbol{\theta})$ is a probability measure on $\mathcal{S}_i$, which is referred to as the \emph{signal structure} or \emph{likelihood function} of agent $i$. The private signals for agent $i$ are generated according to $\ell_i(\mathord{\cdot}\mid\boldsymbol{\theta})$. Let $\Omega$ be an infinite product space encompassing the state $\boldsymbol{\theta}$ and the private signals $\{\mathbf{s}_{i,t}, i \in [n], t\in\mathbb{N}_0\}$. Let $\mathbb{P}\{\mathord{\cdot}\}$ be the probability measure on $\Omega$ with the corresponding expectation operator $\mathbb{E}\{\mathord{\cdot}\}$,  assigning probabilities consistently with the distribution $\nu(\mathord{\cdot})$ and the likelihood functions $\ell_i(\mathord{\cdot}\mid\boldsymbol{\theta})$, $i \in [n]$. Conditioned on $\boldsymbol{\theta}$, the random variables $\{\mathbf{s}_{i,t}$ $,i\in[n]$, $t\in\mathbb{W}\}$ are independent. Consequently, the privately observed signals are independent and identically distributed over time and independent across the agents.

\vspace*{-12pt}
\paragraph*{{\bf Beliefs.}} We let ${\boldsymbol\mu}_{i,t}(\mathord{\cdot})$ be a probability distribution on the set $\Theta$ representing the \emph{opinion} or \emph{belief} at time $t$ of agent $i$ about the realized value of $\boldsymbol{\theta}$. The goal is to study asymptotic learning, i.e. for each agent to learn the true realized value of $\boldsymbol{\theta}$ asymptotically; we denote this truth by $\theta \in \Theta$. Hence, learning amounts to having $\boldsymbol\mu_{i,t}(\mathord{\cdot})$ converge to a point mass centered at $\theta$, where the convergence could be in probability or in the stronger almost sure sense that we use in this work.

At $t= 0$ the value $\boldsymbol{\theta} = \theta$ is selected by nature. Followed by that, $\mathbf{s}_{i,0}$ for each $i\in [n]$ is realized and observed by agent $i$. Then the agent forms an initial Bayesian opinion ${\boldsymbol\mu}_{i,0}(\mathord{\cdot})$ about the value of $\theta$. Given $\mathbf{s}_{i,0}$, and using Bayes rule for each agent $i\in[n]$, the initial belief in terms of the observed signal $\mathbf{s}_{i,0}$ is given by:
\begin{equation}
{\boldsymbol\mu}_{i,0}(\check{\theta}) = \frac{ \nu_i(\check{\theta})\ell_i(\mathbf{s}_{i,0}\mid \check{\theta} )} {\displaystyle\sum_{\hat{\theta} \in \Theta} \nu_i(\hat{\theta})\ell_i(\mathbf{s}_{i,0} \mid \hat{\theta} )},
\label{eq:bayes1}
\end{equation} where $\nu_i(\mathord{\cdot})\in\Delta\Theta$ is an initial full-support ($\nu_i(\hat{\theta})>0$, $\forall \hat{\theta} \in \Theta$) prior belief associated with agent $i$; it represents the subjective biases of agnet $i$ even before making any observations. In particular, for an unbiased agent we have that $\nu(\hat{\theta}) = 1/|\Theta|$, $\forall \hat{\theta}\in\Theta$.

If we assume that each agent $i$ knows the initial priors of her neighbors: $\nu_j(\mathord{\cdot})$, $j\in\mathcal{N}(i)$,\footnote{The assumption of (common) knowledge of priors in the case of rational (Bayesian) agents can be justified as follows: given the same observations of an agent $j$ and in the absence of any past observations or additional data, any agent $i$ should make the same (rational) inference; in the sense that starting form the same belief about the unknown, their updated beliefs given the same observations would be the same, or in Aumann's words, \emph{rational agents cannot agree to disagree}, \cite{aumann1976agreeing}.} then Bayes rule can be exploited to derive the refined opinion ${\boldsymbol\mu}_{i,1}(\mathord{\cdot})$ after agent $i$ observes her neighbors' initial beliefs $\{{\boldsymbol\mu}_{j,0}(\mathord{\cdot}); j \in \mathcal{N}(i)\}$ given by \eqref{eq:bayes1}. Following \cite{rahimian2014non} this leads to
\begin{align}
{\boldsymbol\mu}_{i,1}(\check{\theta}) = \frac{\nu_i(\check{\theta})l_i(\mathbf{s}_{i,0}\mid\check{\theta})\left(\prod_{j\in\mathcal{N}(i)}\frac{{\boldsymbol\mu}_{j,0}(\check{\theta})}{{\nu_j(\check{\theta})}}\right)}{\sum_{\hat{\theta} \in \Theta}\nu_i(\hat{\theta})l_i(\mathbf{s}_{i,0}\mid\hat{\theta})\left(\prod_{j\in\mathcal{N}(i)}\frac{{\boldsymbol\mu}_{j,0}(\hat{\theta})}{{\nu_j(\hat{\theta})}}\right)}.  \label{eq:originalBayesian} 
\end{align} 
How should the belief be updated in the next time steps? A Bayesian agent who cannot recall how the neighbors' beliefs are updated
might instead imitate the preceding structure for all the following time steps $t > 1$. Indeed, in \eqref{eq:originalBayesian} we can substitute $\mathbf{s}_{i,t}$, ${\boldsymbol\mu}_{i,t}(\mathord{\cdot})$, and ${\boldsymbol\mu}_{j,t-1}(\mathord{\cdot})$ for $\mathbf{s}_{i,0}$, ${\boldsymbol\mu}_{i,1}(\mathord{\cdot})$, ${\boldsymbol\mu}_{j,0}(\mathord{\cdot})$. How should the agent update the priors? We study various choices for $\nu_j(\mathord{\cdot}),\forall j$ .

To proceed, for each agent $i$ we also replace $\nu_j(\mathord{\cdot}),\forall j$ with a  time-varying distribution ${\boldsymbol\xi}_{i,j}(\mathord{\cdot},t)$, and argue that a \emph{Rational but Memoryless} agent $i$ would make her rational inference about the opinion ${\boldsymbol\mu}_{j,t-1}(\mathord{\cdot})$ that agent $j$ reports to her at time $t$ according to some  time-varying prior ${\boldsymbol\xi}_{i,j}(\mathord{\cdot},t), j \in [n]$. Here, any choice of distributions ${\boldsymbol\xi}_{i,j}(\mathord{\cdot},t), j \in [n]$ should satisfy the information constraints of  a \emph{Rational but Memoryless} agent, so long as such a choice does not require an agent to recall any information other than what she has just observed $\mathbf{s}_{i,t}$ and what her neighbors have just reported to her ${\boldsymbol\mu}_{j,t-1}(\mathord{\cdot}),j\in\mathcal{N}(i)$. A memoryless yet rational agent of this type can process the beliefs of her neighbors, but cannot recall how these beliefs were formed.
 Accordingly, \eqref{eq:originalBayesian} becomes
\begin{align}
&{\boldsymbol\mu}_{i,t}(\check{\theta}) =  \label{eq:bayesianwithSignalsMAIN3} \\ &\frac{{\boldsymbol\xi}_{i,i}(\check{\theta},t)l_i(\mathbf{s}_{i,t}\mid\check{\theta})\left(\prod_{j\in\mathcal{N}(i)}\frac{{\boldsymbol\mu}_{j,t-1}(\check{\theta})}{{{\boldsymbol\xi}_{i,j}(\check{\theta},t)}}\right)}{\sum_{\hat{\theta} \in \Theta}{\boldsymbol\xi}_{i,i}(\hat{\theta},t)l_i(\mathbf{s}_{i,t}\mid\hat{\theta})\left(\prod_{j\in\mathcal{N}(i)}\frac{{\boldsymbol\mu}_{j,t-1}(\hat{\theta})}{{{\boldsymbol\xi}_{i,j}(\hat{\theta},t)}}\right)}, 
\end{align} for all $\check{\theta} \in \Theta$ and at any $t > 1$. In writing \eqref{eq:bayesianwithSignalsMAIN3}, the time one update is regarded as a function that maps the priors, the private signal, and the neighbors' beliefs to the agent's posterior belief; and in using the time one update in the subsequent steps as in \eqref{eq:bayesianwithSignalsMAIN3}, every time agent $i$ regards each of her neighbors $j\in\mathcal{N}(i)$ as having started from some prior belief ${\boldsymbol\xi}_{i,j}(\mathord{\cdot},t)$ and arrived at their currently reported belief ${\boldsymbol\mu}_{j,t-1}(\mathord{\cdot})$ directly after observing a private signal, hence rejecting any possibility of a past history. Such a rule is of course not the optimum Bayesian update of agent $i$'s belief at any step $t>1$, because the agent is not taking into account the complete observed history of her private signals and neighboring beliefs and is instead, basing her inference entirely on the immediately observed signal and neighboring beliefs; hence, the name \emph{memoryless}. In the next section, we address the choice of random and time-varying priors ${\boldsymbol\xi}_{i,j}(\mathord{\cdot},t), j \in \mathcal{N}(i)$ while examining the properties of convergence and learning under the update rules in \eqref{eq:bayesianwithSignalsMAIN3}.

\section{Asymptotic Analysis of Convergence \& Learning with Log-Linear Updates}\label{sec:convergence}

We begin by forming the $\log$-ratios of beliefs, signals, and priors under the true and false states as ${\boldsymbol\phi}_{i,t}(\check{\theta}) := \log({{\boldsymbol\mu}_{i,t}(\check{\theta})}/{{\boldsymbol\mu}_{i,t}({\theta})})$, $\boldsymbol\lambda_{i,t}(\check{\theta}) $ $:= $ $\log \left({\ell_i(\mathbf{s}_{i,t}|\check{\theta})}/{\ell_i(\mathbf{s}_{i,t}|{\theta})}\right)$, and $\boldsymbol\gamma_{i,j}(\check{\theta},t) $ $:= $ $\log({{\boldsymbol\xi}_{i,j}(\check{\theta},t)}/{{\boldsymbol\xi}_{i,j}({\theta},t)})$ for all $i$, $j$ and $t$. Consequently, \eqref{eq:bayesianwithSignalsMAIN3} can be linearized as follows:
\begin{align}
\boldsymbol\phi_{i,t}(\check{\theta}) = & \, \boldsymbol\gamma_{i,i}(\check{\theta},t) + \boldsymbol\lambda_{i,t}(\check{\theta}) \label{RATIOlog} \\ & + \sum_{j \in \mathcal{N}(i)}(\boldsymbol\phi_{j,t-1}(\check{\theta}) - \boldsymbol\gamma_{i,j}(\check{\theta},t)).
\end{align} The network graph structure is encoded by its adjacency matrix $A$ defined as $[A]_{ij} = 1 \iff (j,i) \in \mathcal{E}$, and $[A]_{ij} = 0$ otherwise. For a strongly connected $\mathcal{G}$ the Perron-Frobenius theory, cf. \cite[Theorem 1.5]{seneta2006non}, implies that $A$ has a simple positive real eigenvalue, denoted by $\rho>0$, which is equal to its spectral radius. Moreover, the left eigenspace associated with $\rho$ is one-dimensional with the corresponding eigenvector $\overline{\alpha} = (\alpha_1,\ldots,\alpha_n)^T$, uniquely satisfying $\sum^{n}_{i=1} \alpha_i = 1$, $\alpha_i>0$, $\forall i \in [n]$, and $\overline{\alpha}^{^T} A = \rho \overline{\alpha}^{^T}$. The quantity $\alpha_i$ is also known as the eigenvector centrality of vertex $i$ in the network, cf. \cite[Section 7.2]{newman2010networks}. Multiplying both sides of \eqref{RATIOlog} by $\alpha_i$ and summing over all $i$ we obtain that 
\begin{align}
\boldsymbol\Phi_{t}(\check{\theta}) = & \, \mbox{tr}\{\boldsymbol\Xi_{t}(\check{\theta})\} + \boldsymbol\Lambda_{t}(\check{\theta}) \label{RATIOlogSUMMED} \\ & + \rho \boldsymbol\Phi_{t-1}(\check{\theta}) - \mbox{tr}\{\boldsymbol\Xi_{t}(\check{\theta}) A^{T}\} \\  = & \sum_{\tau = 0}^{t}\rho^{\tau}\left(\boldsymbol\Lambda_{t-\tau} + \mbox{tr}\{(I - A^{T})\boldsymbol\Xi_{t-\tau}(\check{\theta})\}  \right)
\end{align} where $\boldsymbol\Phi_{t}(\check{\theta}) := \sum_{i=1}^{n} \alpha_i \boldsymbol\phi_{i,t}(\check{\theta})$ and $\boldsymbol\Lambda_{t}(\check{\theta}) := \sum_{i=1}^{n} \alpha_i \boldsymbol\lambda_{i,t}(\check{\theta})$ are global (network-wide) random variables, and $\boldsymbol\Xi_{t}(\check{\theta})$ is a random $n \times n$ matrix whose $i,j$-th entry is given by $[\boldsymbol\Xi_{t}(\check{\theta})]_{i,j} = \alpha_i \boldsymbol\gamma_{i,j}(\check{\theta},t)$. At each epoch of time, $\boldsymbol\Phi_{t}(\check{\theta})$ characterizes how biased (away from the truth and towards $\check{\theta}$) the network beliefs are, and $\boldsymbol\Lambda_{t}(\check{\theta})$ measures the information content of the received signal across all the agents in the network. Note that in writing \eqref{RATIOlogSUMMED}, we use the fact that
\begin{align}
\sum_{i=1}^{n}\alpha_i\sum_{j \in \mathcal{N}(i)}\boldsymbol\phi_{j,t-1}(\check{\theta}) = \overline{\alpha}^{^T} A \overline{\boldsymbol\phi}_{t-1}(\check{\theta}) \\ = \rho \overline{\alpha}^{^T}\overline{\boldsymbol\phi}_{t-1}(\check{\theta}) = \rho\boldsymbol\Phi_{t-1}(\check{\theta}),\label{eq:GAMA}
\end{align} where $\overline{\boldsymbol\phi}_{t}(\check{\theta})$ $:=$ $(\boldsymbol\phi_{1,t}(\check{\theta}),\ldots,\boldsymbol\phi_{n,t}(\check{\theta}))^{T}$. On the other hand, since the received signal vectors $\{\mathbf{s}_{i,t}, i \in [n] , t\in\mathbb{N}_{0} \}$ are i.i.d. over time, $\{\boldsymbol\Lambda_{t}(\check{\theta}),t\in\mathbb{N}_{0}\}$ constitutes a sequence of i.i.d. random variables satisfying
\begin{align}
& \mathbb{E}\left\{\boldsymbol\Lambda_{t}(\check{\theta})\right\}  =  \displaystyle\sum_{i=1}^{n} \alpha_i \mathbb{E}\{\boldsymbol\lambda_{i,t}(\check{\theta})\}   =  \displaystyle\sum_{i=1}^{n} \alpha_i {\lambda}_i(\check{\theta}) \leqslant 0,\label{eq:RATElambda} 
\end{align} where $\overline{\lambda}(\check{\theta}) := (\lambda_1(\check{\theta}),\ldots,\lambda_n(\check{\theta}))^{T}:= $
\begin{align}
&-\left(D_{KL}\left(\ell_1( \mathord{\cdot} |\theta) || \ell_1( \mathord{\cdot} |\check{\theta}) \right), \ldots, D_{KL}\left(\ell_n( \mathord{\cdot} |\theta) || \ell_n( \mathord{\cdot} |\check{\theta}) \right) \right)^{T},
\end{align} and the non-positivity of \eqref{eq:RATElambda} follows from the information inequality for the Kullback-Leibler divergence: $D_{KL}\left(\mathord{\cdot}|| \mathord{\cdot} \right) \geq 0$, and is strict whenever $\ell_i( \mathord{\cdot} |\check{\theta}) \not\equiv \ell_i( \mathord{\cdot} |\theta)$ for some $i$, i.e. $\exists s \in \mathcal{S}_i, i\in [n]$ such that $\ell_i( s |\check{\theta}) \neq \ell_i( s |\theta)$, cf. \cite[Theorem 2.6.3]{cover2006elements}. In particular, if for all $\check{\theta}\neq\theta$ there exists an agent $i$ with  ${\lambda}_i(\check{\theta}) < 0$, then $\mathbb{E}\left\{\boldsymbol\Lambda_{t}(\check{\theta})\right\} < 0$ and we say that the truth $\theta$ is \emph{globally identifiable}. Indeed, if any agent is to learn the truth, then we need that $\boldsymbol{\Phi}_t(\check{\theta}) \to -\infty$ as $t\to \infty$ for all the false states $\check{\theta} \neq \theta$. However, in a strongly connected graph every node has a degree greater than or equal to one so that $\rho \geq 1$, \cite[Chapter 2]{richard2011mutually}. If $\rho > 1$, then the term $\rho^{t}\boldsymbol\Lambda_{0}(\check{\theta})$ increases in variance as $t \to \infty$, and unless $\boldsymbol\Lambda_{0}(\check{\theta})<\epsilon$ with $\mathbb{P}$-probability one for some $\epsilon<0$, almost sure convergence to $-\infty$ for $\boldsymbol\Phi_{t}(\check{\theta})$ in \eqref{RATIOlogSUMMED} cannot hold true. In a directed circle where $\rho = 1$, one may take $\boldsymbol\xi_{i,j}(\mathord{\cdot},t) \equiv \nu_j(\mathord{\cdot}) \equiv \nu(\mathord{\cdot})$ for all $i$, $j$ and $t$. Consequently, at each epoch of time each agent $i$ assumes that all her neighbors have started from the initial common prior belief $\nu(\mathord{\cdot})$ and have arrived at their current beliefs directly, thus forgetting any history of observed signals and exchanged opinions. The geometric progression in \eqref{RATIOlogSUMMED} then reduces to sum of i.i.d. variables in $\mathcal{L}^1$; and by the strong law of large numbers, cf. \cite[Section X.1]{feller1971introduction}, it converges almost surely to the mean value; hence, $ \boldsymbol\Phi_{t}(\check{\theta})$ $ =$ $ {\beta}(\check{\theta}) + \sum_{\tau=0}^{t}\boldsymbol\Lambda_{\tau}(\check{\theta})$ $ \to$ $ {\beta}(\check{\theta}) + (t+1)\mathbb{E}\left\{\boldsymbol\Lambda_{0}(\check{\theta})\right\}$ $ \to$ $ -\infty$, as $t \to \infty$, provided that $\mathbb{E}\left\{\boldsymbol\Lambda_{0}(\check{\theta})\right\} < 0$, i.e. the truth is globally identifiable. Here ${\beta}(\check{\theta}):=\overline{\alpha}^{^{T}} \overline{\psi}(\check{\theta})$, where $\overline{\psi}(\check{\theta})$ $ :=$ $ \left({\psi}_1(\check{\theta}),\ldots, {\psi}_n(\check{\theta})\right)^{T}$ is the stacked vector of initial prior $\log$-ratios with  ${\psi}_i(\check{\theta}):=\log(\nu_i(\check{\theta})/\nu_i({\theta}))$, $\forall i\in[n]$. Thus as defined, ${\beta}(\check{\theta})$ measures the network-wide bias  in the agents' initial priors toward the state $\check{\theta}$. Accordingly,  \eqref{eq:bayesianwithSignalsMAIN3} for a circular network with common priors becomes
\begin{align}
{\boldsymbol\mu}_{i,t}(\hat{\theta}) = \frac{ {\boldsymbol\mu}_{j,t-1}(\hat{\theta})\ell_i(\mathbf{s}_{i,t}\mid\hat{\theta})} {\displaystyle\sum_{\hat{\theta} \in \Theta}{\boldsymbol\mu}_{j,t-1}(\hat{\theta})\ell_i(\mathbf{s}_{i,t}\mid\hat{\theta})}, \forall \hat{\theta} \in \Theta,
\label{eq:bayesSINGLE_neighborReplaced}
\end{align} where $j\in [n]$ is the unique vertex $j \in \mathcal{N}(i)$. The above is the same as Bayesian update of a single agent (with no neighbors to communicate with), except that the self belief $\boldsymbol\mu_{i,t-1}(\mathord{\cdot})$ in the right-hand side is replaced by the belief $\boldsymbol\mu_{j,t-1}(\mathord{\cdot})$ of the unique neighbor $\{j\} = \mathcal{N}(i)$ in the circle. 

\cite{LWCcircleTree} show that under \eqref{eq:bayesSINGLE_neighborReplaced} the agents in a directed circle learn the truth asympototically exponentially fast and at the rate $\min_{\check{\theta}\neq\theta}$ $(-1/n)\sum_{i=1}^{n} {\lambda}_i(\check{\theta})$, so that $
\lim_{t\to\infty}$ $\frac{1}{t}$ $ {\boldsymbol\phi}_{i,t}(\check{\theta})$ $ = $ $(1/n)$ $\sum_{i=1}^{n} {\lambda}_i(\check{\theta})$, almost surely, for all $\check{\theta} \neq \theta$. \cite{LWCrandomWalk} show the application of the update rule in \eqref{eq:bayesSINGLE_neighborReplaced} to general strongly connected topologies (no necessarily circular), where agents have more than just a single neighbor in their neighborhoods. Accordingly, at every step of time agent $i$ chooses a neighbor $j \in \mathcal{N}(i)$  independently at random and applies \eqref{eq:bayesSINGLE_neighborReplaced} with the reported belief of the chosen neighbor. The probabilities for the choice of neighbors at every point in time are given by a row stochastic matrix $P$, with entries $[P]_{ij}>0$ for every $j\in \mathcal{N}(i)$. Each entry $[P]_{ij}$ is the probability for neighbor $j$ being chosen by agent $i$ at any point in time. Subsequently, we can show that the belief ratio for agent $i$ at every time $t$ is given as the sum of log-likelihood ratios of private signals of various agents across the network and at times $0$ to $t$. The choice of agent at every time is given by a random walk $\mathbf{i}_\tau$, $\tau\in[t]$ that starts from node $\mathbf{i}_0 = i$ at time $t$ and proceeds in the reversed time direction, eventually terminating at some node $\mathbf{i}_t$. Accordingly, the log-belief ratio of agent $i$ at any time $t$ can be expressed as
${\boldsymbol\phi}_{i,t}(\check{\theta})$ $ = $ $ \psi_{\mathbf{i}_t}(\check{\theta})$ $ +$ $ \sum^{t}_{\tau=0} \boldsymbol\lambda_{{\mathbf{i}_\tau},t-\tau}(\check{\theta})$. Next note from the ergodic theorem, cf. \cite[Theorem 1.10.2]{norris1999markov}, that the average time spent in any state $m\in[n]$ converges almost surely to its stationary probability $\pi_m$ associated with the probability transition matrix $P$, and this together with the strong law yields $\lim_{t\to\infty}$ $\frac{1}{t}$ $ {\boldsymbol\phi}_{i,t}(\check{\theta}) $ $=$ $ \sum_{i=1}^{n}$ $ \pi_i {\lambda}_i(\check{\theta})$,  almost surely for all $\check{\theta}\neq\theta$. Hence if the truth is globally identifiable, then in a strongly connected network every agent learns the truth at an asymptotic rate that is exponentially fast and is expressed above, as the sum of the relative entropies between the signal structures of every agent weighted by the stationary distribution of the random walk, which recovers the same asymptotic rate for the update proposed by \cite{JaMoTa13}.

Fixing the priors over time will not result in convergence of beliefs, except in very specific cases as discussed above. In the sequel, we investigate the properties of convergence and learning under the update rules in \eqref{eq:bayesianwithSignalsMAIN3}, where the parameterizing priors ${\boldsymbol\xi}_{i,j}(\mathord{\cdot},t)$ are chosen to be random and time-varying variables, leading to the log-linear updating of the agents' beliefs over time. We distinguish two cases depending on whether the agents do not recall their own self-beliefs or they do, leading respectively to time-invariant or time-varying log-linear update rules.


\subsection{Priors Set to a Geometric Average} \label{sec:loglinconst} 


It is notable that the memoryless Bayesian update in \eqref{eq:bayesianwithSignalsMAIN3} has a log-linear structure similar to the Non-Bayesian update rules studied by \cite{kamyar_CDC_2010,shahin_CDC_2013,nedic2014nonasymptotic,6874893,bandyopadhyay2014distributed}; and the roots for such a geometric averaging of the neighboring beliefs can be traced to logarithmic opinion pools as studied by \cite{gilardoni1993reaching} and \cite{rufo2012log}. Motivated by this analogy, we propose setting the time-varying priors $\boldsymbol\xi_{i,j}(\mathord{\cdot},t), j\in \mathcal{N}(i)$ of each agent $i$ and at every time $t$, proportionally to the geometric average of the beliefs reported to her by all her neighbors at every time $t$: $\prod_{j\in\mathcal{N}(i)}{{\boldsymbol\mu}_{j,t-1}(\mathord{\cdot})}^{1/d(i)}$. Therefore, \eqref{eq:bayesianwithSignalsMAIN3} becomes 
\vspace{-3mm}
\begin{align}
{\boldsymbol\mu}_{i,t}(\check{\theta}) = \frac{ l_i(\mathbf{s}_{i,t}\mid\check{\theta}) \left(\prod_{j\in\mathcal{N}(i)}{{\boldsymbol\mu}_{j,t-1}(\check{\theta})}\right)^{1/d(i)}}{\sum_{\hat{\theta} \in \Theta} l_i(\mathbf{s}_{i,t}\mid\hat{\theta}) \left(\prod_{j\in\mathcal{N}(i)}{{\boldsymbol\mu}_{j,t-1}(\hat{\theta})}\right)^{1/d(i)}}.\label{eq:bayesianwithSignalsMAIN33}
\end{align}
To analyze the evolution of beliefs with this choice of priors, let $\overline{\boldsymbol\lambda}_{t}(\check{\theta})$ $:=$ $(\boldsymbol\lambda_{1,t}(\check{\theta}),\ldots,\boldsymbol\lambda_{n,t}(\check{\theta}))^{T}$ be the stacked vector of $\log$-likelihood ratios of received signals for all agents at time $t$. Hence, we can write the vectorized update $\overline{\boldsymbol\phi}_{t}(\check{\theta}) = T \overline{\boldsymbol\phi}_{t-1}(\check{\theta}) + \overline{\boldsymbol\lambda}_{t}(\check{\theta})$, where $T$ is the normalized adjacency of the graph defined by $[T]_{ij} = \frac{1}{d(i)}[A]_{ij}$ for all $i$ and $j$. We can now iterate the vectorized update to get $ \overline{\boldsymbol\phi}_{t}(\check{\theta})$ $ =$ $ \sum_{\tau=0}^{t} T^{\tau} \overline{\boldsymbol\lambda}_{t-\tau}(\check{\theta}) + T^{t} \overline{\psi}(\check{\theta})$. Next note from the analysis of convergence for DeGroot model, cf. \cite[Proporition 1]{GolubWisdomCrowd}, that for a strongly connected network $\mathcal{G}$ if it is aperiodic (meaning that one is the greatest common divisor of the lengths of all its circles), then $\lim_{\tau\to\infty}T^{\tau} = \mathds{1}\overline{s}^{^{T}}$, where $\overline{s}:=(s_1,\ldots,s_n)^{T}$ is the unique left eigenvector associated with the unit eigenvalue of $T$ and satisfying $\sum_{i=1}^{n}s_i = 1$, $s_i>0$, $\forall i$. Hence, the Ces\`{a}ro mean  together with the strong law implies that $ \lim_{t\to\infty}$ $\frac{1}{t}$ $\boldsymbol{\phi}_{i,t}(\check{\theta})$ $ =$ $ -\sum_{i=1}^{n}$ $ s_i {\lambda}_i(\check{\theta})$, almost surely for all $\check{\theta} \neq \theta$, and the agents learn the truth asymptotically exponentially fast, at the rate $
\min_{\check{\theta}\neq\theta}$ $\sum_{i=1}^{n} -s_i  {\lambda}_i(\check{\theta})$.


\subsection{Agents who Recall Their Self Beliefs} \label{sec:timevaryingloglinear}

If the agents $i\in[n]$ recall their self-beliefs $\boldsymbol\mu_{i,t-1}(\mathord{\cdot}), i\in[n]$ when making decisions or performing inferences at time $t$, then we set $\boldsymbol\xi_{i,i}(\mathord{\cdot},t) \equiv \boldsymbol\mu_{i,t-1}(\mathord{\cdot})$ for all $i$ and $t$. Furthermore, we set $\boldsymbol\xi_{i,j}(\mathord{\cdot},t) \equiv \boldsymbol\mu_{j,t-1}(\mathord{\cdot})^{\eta_t}/\boldsymbol\zeta_{j}(t)$ for all $i$, $j \in \mathcal{N}(i)$ and $t$, where $\boldsymbol\zeta_{j}(t) := \sum_{\hat{\theta}\in\Theta}\boldsymbol\mu_{j,t-1}(\hat{\theta})^{\eta_t}$ is the normalization constant to make the exponentiated probabilities sum to one. The choice of $0<\eta_t<1$ as time-varying exponents to be determined shortly, is motivated by the requirements of convergence under \eqref{eq:bayesianwithSignalsMAIN3}. Subsequently, we investigate the following log-linear update rule with time-varying coefficients
\begin{align}
& {\boldsymbol\phi}_{i,t}(\check{\theta}) = {\boldsymbol\phi}_{i,t-1}(\check{\theta}) + \boldsymbol\lambda_{i,t}(\check{\theta}) + (1-\eta_t)\sum_{j\in\mathcal{N}(i)} {\boldsymbol\phi}_{j,t-1}(\check{\theta}),
\label{eq:loglinearUpdate}
\end{align} where $1-\eta_t$ is the weight that the agent puts on her neighboring beliefs (relative to her own) at any time $t$. Using $ x_t:= \rho(1-\eta_t)$ and $B := (1 / \rho) A $, the previous equation can be written in vectorized format as follows
\begin{align}
 \overline{\boldsymbol\phi}_t(\check{\theta}) = &  (I + x_t B)\overline{\boldsymbol\phi}_{t-1}(\check{\theta}) + \overline{\boldsymbol\lambda}_{t}(\check{\theta})\\
  = & \sum^{t}_{\tau = 0} P^{(t,\tau)} \overline{\boldsymbol{\lambda}}_{\tau}(\check{\theta}) + P^{(t,0)} \overline{\psi}(\check{\theta}),
\label{eq:loglinearUpdateBeliefRatiosMatrixForm0} 
\end{align} where $P^{(t,t)}:= I$, and $P^{(t,\tau)}:=\prod^{t}_{u=\tau+1}(I + x_{u}B)$ for $\tau < t$. Next note that $P^{(t,\tau)}$ can be expanded as follows
\begin{align}
P^{(t,\tau)}  =\sum^{t-\tau}_{j=0}M_j^{(t,\tau)}B^{j},
 \label{eq:loglinearUpdateBeliefRatiosMatrixForm2} 
\end{align} where $M_{0}^{(t,\tau)} = 1$, $\tau\leq t$ and 
\begin{align}
M_{j}^{(t,\tau)} = \sum^{\scriptstyle t-j+1}_{\scriptstyle u_1 = \tau+1}\sum^{\scriptstyle t-j+2}_{\scriptstyle u_2 = u_1 + 1}\ldots\sum^{\scriptstyle t}_{\scriptstyle u_{j} = u_{j-1} + 1}x_{u_1}x_{u_2}\ldots x_{u_{j}}.
\end{align} Consequently, \eqref{eq:loglinearUpdateBeliefRatiosMatrixForm0} can be rewritten as   
\begin{align}
& \overline{\boldsymbol\phi}_t(\check{\theta}) = \label{eq:loglinearUpdateBeliefRatiosMatrixForm00}  \\ & \sum^{t}_{\tau = 0} \sum^{t-\tau}_{j=0}M_j^{(t,\tau)}B^{j} \overline{\boldsymbol{\lambda}}_{\tau}(\check{\theta}) + \sum^{t}_{j=0}M_j^{(t,0)}B^{j} \overline{\psi}(\check{\theta}),
\end{align}
To proceed, for fixed $\tau$ and $j$ let $M_{j}^{(\tau)} := \lim_{t\to\infty}M_{j}^{(t,\tau)}$. Next consider the summands in \eqref{eq:loglinearUpdateBeliefRatiosMatrixForm00} for each $\tau$, $0 \leq \tau \leq t$. Note that for $j$ fixed, $\{B^{j} \overline{\boldsymbol{\lambda}}_{\tau}(\check{\theta}), \tau \in \mathbb{N}_{0} \}$ is a sequence of independent and identically distributed random vectors. On the other hand, since $B$ has unit spectral radius and given that $x_{u}>0$, having $M_{1}^{(0)}:=\sum_{u=1}^{\infty}x_{u} <\infty$ is sufficient to ensure that the random vectors $M_{j}^{(t,\tau)} B^{j} \overline{\boldsymbol\lambda}_{\tau}(\check{\theta})$ are all in $\mathcal{L}^2$ and have variances that are  bounded uniformly in the choice of $t$, $\tau$. This is because for any $t$, $\tau$, and $j$ we have that 
\begin{align}
M_{j}^{(t,\tau)} \leq  M_{j}^{(\tau)}  \leq M_{j}^{(0)}   \leq   \frac{1}{j!}\left(M_{1}^{(0)}\right)^{j}, 
\end{align} all as a consequence of positivity, $x_u > 0$. In particular, with $M_{1}^{(0)} <\infty$ we can bound  
\begin{align}
\left\lVert\sum^{\infty}_{j=0}M_j^{(0)}B^{j} \overline{\psi}(\check{\theta})\right\rVert  &\leq  \sum^{\infty}_{j=0}  M_j^{(0)}  \left\lVert B^{j} \overline{\psi}(\check{\theta})\right\rVert \\ & \leq \left\lVert \overline{\psi}(\check{\theta})\right\rVert  \exp(M_{1}^{0}), \label{eq:boundedBIAS}
\end{align}  so that the contribution made by the initial bias of the network is asymptotically bounded and therefore sub-dominant when 
\begin{align}
\lim_{t\to\infty} \sum^{t}_{\tau = 0} \sum^{t-\tau}_{j=0}M_j^{(t,\tau)}B^{j} \overline{\boldsymbol{\lambda}}_{\tau}(\check{\theta}) = (-\infty)_{_n}, 
\end{align} almost surely; here, by $(-\infty)_n$ we mean the entry-wise convergence of the column vector to $-\infty$ for the each of the $n$ entries, corresponding to the $n$ agents. In the sequel we investigate conditions under which this almost sure convergence would hold true.

We begin by noting that the condition $M_{1}^{(0)} <\infty$ is indeed necessary for convergence, because if $M_{1}^{(0)} = \infty$, then the term $M_{1}^{(t,0)} B \overline{\boldsymbol\lambda}_{0}(\check{\theta})$ appearing in \eqref{eq:loglinearUpdateBeliefRatiosMatrixForm00} for $j=1$ and $\tau = 0$ increases unbounded in its variance as $t \to \infty$, so that \eqref{eq:loglinearUpdateBeliefRatiosMatrixForm00} cannot converge in an almost sure sense.

Next note that with $M_{1}^{(0)} <\infty$ we can invoke Kolmogorov's criterion, \cite[Section X.7]{feller1971introduction}, to get that as $t \to \infty$ the summation in \eqref{eq:loglinearUpdateBeliefRatiosMatrixForm00} converges almost surely to its expected value, i.e. the following almost sure limit holds true
\begin{align}
&\lim_{t\to\infty}{\boldsymbol\phi}_{i,t}(\check{\theta}) = \label{agent_sum} \\ &\sum^{\infty}_{\tau = 0} \sum^{\infty}_{j=0}M_j^{(\tau)}\left[B^{j} \overline{\lambda}(\check{\theta})\right]_i  + \sum^{\infty}_{j=0}M_j^{(0)}\left[B^{j} \overline{\psi}(\check{\theta})\right]_i, 
\end{align} the second term being bounded per \eqref{eq:boundedBIAS}. Moreover, for a strongly connected social network $\mathcal{G}$, if it is aperiodic, then the matrix $B$ is a primitive matrix; and in particular for all $j\geq d:=\mbox{diam}(\mathcal{G})+1$, every entry of $B^{j}$ is strictly greater than zero, and if the truth is globally identifiable then one can take an absolute constant $\epsilon>0$ such that $\left[B^{j} \overline{\lambda}(\check{\theta})\right]_i<-\epsilon$ whenever $j\geq d$, while $\left[B^{j} \overline{\lambda}(\check{\theta})\right]_i \leq 0$ for any $j$. Subsequently, we get that
\begin{align}
\sum^{\infty}_{\tau = 0} \sum^{\infty}_{j=0} M_j^{(\tau)}\left[B^{j} \overline{\lambda}(\check{\theta})\right]_i &\leq -\epsilon \sum^{\infty}_{\tau = 0} \sum^{\infty}_{j=d}M_j^{(\tau)}. \label{eq:upperbound}  
\end{align} Combining the results of \eqref{eq:boundedBIAS}, \eqref{agent_sum} and \eqref{eq:upperbound}  leads to the following characterization: {all agents will learn the truth (that is $\lim_{t\to\infty}{\boldsymbol\phi}_{i,t}(\check{\theta}) = -\infty$, almost surely for all $i$ and any $\check{\theta}\neq\theta$), if $ M^{(0)}_{1}<\infty$ and $\sum^{\infty}_{\tau = 0} \sum^{\infty}_{j=d}M_j^{(\tau)} = \infty$. Notice the preceding conditions are indeed not far from necessity. Firstly, we need $ M^{(0)}_{1}<\infty$ to bound the growth of variance for convergence, as noted above. Moreover, with $B$ having a unit spectral radius we can bound 
\begin{align}
\left|\left[B^{j} \overline{\lambda}(\check{\theta})\right]_i \right| \leq \left\lVert B^{j} \overline{\lambda}(\check{\theta}) \right\rVert \leq \left\lVert \overline{\lambda}(\check{\theta}) \right\rVert, 
\end{align} so that we can lower-bound $\lim_{t\to\infty}{\boldsymbol\phi}_{i,t}(\check{\theta})$ in \eqref{agent_sum} as follows
\begin{align}
-\left\lVert \overline{\lambda}(\check{\theta}) \right\rVert \sum^{\infty}_{\tau = 0} \sum^{\infty}_{j=0}M_j^{(\tau)} - \left\lVert \overline{\psi}(\check{\theta})\right\rVert  \exp(M_{1}^{0}) \leq  \lim_{t\to\infty}{\boldsymbol\phi}_{i,t}(\check{\theta}).
\end{align} Consequently, if $\sum^{\infty}_{\tau = 0} \sum^{\infty}_{j=0}M_j^{(\tau)} < \infty$,  then $\lim\limits_{t\to\infty}{\boldsymbol\phi}_{i,t}(\check{\theta})$ is almost surely bounded away from $-\infty$ and agents do not learn the truth. 

For a strongly connected and aperiodic social network $\mathcal{G}$, matrix $B$ has a single eigenvalue at one (corresponding to the largest eigenvalue of the adjacency $A$) and all of the other eigenvalues of $B$ have magnitudes strictly less than one. Therefore, by the iterations of the power method, \cite[Section 11.1]{newman2010networks}, we know that $\left[B^j \overline{\lambda}(\check{\theta})\right]_i$ converges to $\sum_{k=1}^{n} \alpha_k {\lambda}_k(\check{\theta})$ for every $i$, and the convergence is geometrically fast in the magnitude-ratio of the first and second largest eigenvalues of the adjacency matrix $A$. Subsequently, for a strongly connected and aperiodic social network $\mathcal{G}$, we can replace $-\epsilon$ and $d$ in \eqref{eq:upperbound} by $\epsilon + \sum_{i=1}^{n} \alpha_i {\lambda}_i(\check{\theta})<0$ and some constant $D(\epsilon)$. Here, $\epsilon>0$ is a small but arbitrary and $D(\epsilon)$ is chosen large enough in accordance with the geometric rate of $\left[B^j \overline{\lambda}(\check{\theta})\right]_i \to \sum_{k=1}^{n} \alpha_k {\lambda}_k(\check{\theta})$, such that $|\left[B^j \overline{\lambda}(\check{\theta})\right]_i - \sum_{k=1}^{n} \alpha_k {\lambda}_k(\check{\theta})|$ $<$ $\epsilon$ for all $j\geq D(\epsilon)$, and the analysis of convergence and its rate can thus be refined.

Furthermore, having 
\begin{align}
K_1 &< \liminf_{t\to\infty}\frac{1}{t} \sum^{t}_{\tau = 0} \sum^{t-\tau}_{j=d}M_j^{(t,\tau)} \\  & \leq  \limsup_{t\to\infty}\frac{1}{t} \sum^{t}_{\tau = 0} \sum^{t-\tau}_{j=d}M_j^{(t,\tau)} < K_2
\end{align} for some positive constants $0 < K_1 < K_2$ implies that the learning rate is asymptotically exponentially fast, as was the case for all the other update rules that we discussed in this paper. However, depending on how slow or fast (compared to $t$) is the convergence $\sum^{t}_{\tau = 0} \sum^{t-\tau}_{j=d}M_j^{(t,\tau)} \to \infty$ as $t\to\infty$, the almost sure asymptotic rate at which for some $\check{\theta}\neq\theta$ and $i\in[n]$, ${\boldsymbol\mu}_{i,t}(\check{\theta}) \to 0$ as $t\to \infty$ could be slower or faster than an exponential.



\vspace{-15pt}

\paragraph*{\bf Time-Invariant Log-Linear Updates with Weighted Self-Beliefs.} For such agents as in Subsection~\ref{sec:timevaryingloglinear} who recall their immediate self-beliefs $\boldsymbol\mu_{i,t}(\mathord{\cdot})$, if we relax the requirement that $\boldsymbol\xi_{i,i}(\mathord{\cdot},t) \equiv \boldsymbol\mu_{i,t-1}(\mathord{\cdot})$ then it is possible to achieve asymptotic almost sure exponentially fast learning using time-invariant updates just as in Subsection \ref{sec:loglinconst}. In particular, for any $0<\eta<1$ fixed, we can set $\boldsymbol\xi_{i,i}(\mathord{\cdot},t)$ proportional to $\boldsymbol\mu_{i,t-1}(\mathord{\cdot})^{\eta}$ for all $i$, and we can further set $\boldsymbol\xi_{i,j}(\mathord{\cdot},t)$  at every time $t$, for any $i$, and all $ j\in \mathcal{N}(i)$ proportional to ${{\boldsymbol\mu}_{j,t-1}(\mathord{\cdot})}^{1 - (1-\eta)/d(i)}$. Subsequently, \eqref{eq:bayesianwithSignalsMAIN3} becomes 
\begin{align}
{\boldsymbol\mu}_{i,t}(\check{\theta}) = \frac{ l_i(\mathbf{s}_{i,t}\mid\check{\theta}) \boldsymbol\mu_{i,t-1}(\check{\theta})^{\eta} \left(\prod_{j\in\mathcal{N}(i)}{{\boldsymbol\mu}_{j,t-1}(\check{\theta})}\right)^{\frac{1-\eta}{d(i)}}} {\sum\limits_{\hat{\theta} \in \Theta} l_i(\mathbf{s}_{i,t}\mid\hat{\theta}) \boldsymbol\mu_{i,t-1}(\hat{\theta})^{\eta} \left(\prod_{j\in\mathcal{N}(i)}{{\boldsymbol\mu}_{j,t-1}(\hat{\theta})}\right)^{\frac{1-\eta}{d(i)}}}.\label{eq:bayesianwithSignalsMAIN333}
\end{align} To analyze \eqref{eq:bayesianwithSignalsMAIN333}, we form the $\log$-belief and likelihood ratios and  set $B = (\eta I + (1-\eta) T)$, where $T$ is the same normalized adjacency as in Subsection \ref{sec:loglinconst}. Hence, we recover the vectorized iterations $\overline{\boldsymbol\phi}_{t}(\check{\theta}) = B \overline{\boldsymbol\phi}_{t-1}(\check{\theta}) + \overline{\boldsymbol\lambda}_{t}(\check{\theta}) =$ $ \sum_{\tau=0}^{t} B^{\tau} \overline{\boldsymbol\lambda}_{t-\tau}(\check{\theta}) + B^{t} \overline{\psi}(\check{\theta})$ and it follows from \cite[Theorems 5.1.1 and  5.1.2]{FMC}  that for a strongly connected network $\mathcal{G}$,  $\lim_{\tau\to\infty}B^{\tau} = \mathds{1}{\overline{s}}^{^{T}}$, where $\overline{s}:=(s_1,\ldots,s_n)^{T}$ is the unique stationary distribution associated with the Markov chain whose probability transition matrix is $T$ (or equivalently $B$); whence it follows from the Ces\`{a}ro mean and the strong law that the agents learn the truth asymptotically exponentially fast, at the rate $
\min_{\check{\theta}\neq\theta}$ $\sum_{i=1}^{n} -s_i  {\lambda}_i(\check{\theta})$, similar to Subsection \ref{sec:loglinconst}. Note that here, unlike Subsection \ref{sec:loglinconst} but similarly to \cite{LWCrandomWalk}, we only use properties of ergodic chains and existence and uniqueness of their stationary distributions; hence, relaxing the requirement for the social network to be aperiodic.

\vspace{-17pt}
\paragraph*{\bf Concluding Remarks.}  Learning without recall is a model of belief aggregation in networks by replicating the rule that maps the initial priors, neighboring beliefs, and private signal to Bayesian posterior at one time step for all future time steps; this way the complexities of a fully rational inference at the forthcoming epochs are avoided, while some essential features of Bayesian inference are preserved. We showed how appropriate choice of priors for the so-called rational but memoryless agents in this model can provide a behavioral foundation for the log-linear structure of some of the non-Bayesian update rules that are studied in the literature. We further investigated the rate and requirements of asymptotic and almost sure learning with these choices of priors. In particular, we identified $\sum_{u=1}^{\infty}x_u < \infty$ as a necessary condition for convergence of beliefs with log-linear time-varying updates, in the case of agents who have no recollection of the past, excepting their own immediate beliefs. With $x_t$ representing the relative weight on the neighboring beliefs, such agents facing individual identification problems can still learn the truth in a strongly connected and aperiodic network by relying on each other's observations; provided that as time evolves they put less and less weight on the neighboring beliefs and rely more on their private observations: $x_t \to 0$ as $t\to\infty$.


\vspace*{-15pt}

\bibliography{BayesRef,refDecentralizedDetection}

\end{document}